# Light Speed Invariance is a Remarkable Illusion

**Stephan J. G. Gift**
**Department of Electrical and Computer Engineering**
**Faculty of Engineering**
**The University of the West Indies**
**St. Augustine, Trinidad, West Indies**

*"Einstein's special theory of relativity requires that the one-way velocity of light be a constant. If that turns out not to be so, special relativity falls."*
**Paul A. LaViolette, Genesis of the Cosmos, 2004.**

*"Any clear sign of a variation in c, the speed of light, as the Earth [revolved] would prove that the aether existed."*
**George Smoot, Wrinkles in Time, 1993.**

*"Those physicists - and they are many - who now regard belief in the possibility of an ether as a superstition have simply not learnt the lessons of history, which teach us that discarded ideas have a way of returning to favour."*
**Herbert Dingle, Science at the Crossroads, 1972.**

**Abstract.** Though many experiments appear to have confirmed the light speed invariance postulate of special relativity theory, this postulate is actually unverified. This paper resolves this issue by first showing the manner in which an illusion of light speed invariance occurs in two-way light speed measurement in the framework of a semi-classical absolute space theory. It then demonstrates a measurable variation of the one-way speed of light, which directly invalidates the invariance postulate and confirms the existence of the preferred reference frame of the absolute space theory.

## 1.    Introduction

A fundamental tenet of Einstein's Special Theory of Relativity is the Light Speed Invariance Postulate according to which the speed of light is constant in all inertial frames [1-4]. This postulate is used to derive the Lorentz Transformations relating the coordinates in different inertial frames and these transformations are in turn used to derive the length contraction and frequency reduction formulae of special relativity.

The light speed invariance postulate has been subjected to numerous tests over the past century [5] as a result of which most scientists believe that it has been confirmed. Thus it is often claimed that the postulate has been verified by the classic experiments of Michelson and Morley [6] and Kennedy and Thorndike [7] and by the many later improved versions of these experiments such as those by Brillet and Hall [8] and Hils and Hall [9]. However while the results of this class of experiments suggest a constant light speed $c$, they do not directly test one-way light speed. In experiments that do attempt a one-way light speed test such as those by Gagnon et.al. [10] and Krisher et.al. [11], Zhang [5] has pointed out that these are not true one-way tests because of the inability to independently synchronize the clocks involved. Thus 100 years after the introduction of the relativistic paradigm, light speed invariance on which the paradigm is based remains an open issue, despite the strange declaration by the scientific community that it is correct by definition! Ives has in fact described it as untenable [12].

One-way light speed testing is therefore necessary in order to determine the validity of the light speed invariance postulate. This test can differentiate between Special Relativity Theory, which involves light speed invariance, and the Maxwell-Lorentz Ether Theory, which involves light speed variation relative to a moving observer. This latter theory in its modern form is a semi-classical Absolute Space Theory in which light propagates isotropically at a speed $c$ in a preferred or absolute reference frame. In such an absolute frame and in keeping with classical analysis, the one-way speed of light changes according to the observer's motion relative to the preferred frame.   The theory incorporates the ether-induced Fitzgerald-Larmor-Lorentz (FLL) contractions experimentally confirmed by Ives [13] according to which a rod of length $l_o$ in a preferred frame, when moving with speed $w$ relative to that preferred frame, is shortened to a length $l$ given by

$$l = l_o (1 - w^2 / c^2)^{1/2} \qquad\qquad\qquad (1.1)$$

and a system of frequency $f_o$ when stationary in the preferred frame, has a reduced frequency $f$ given by

$$f = f_o(1 - w^2/c^2)^{1/2} \qquad (1.2)$$

This incorporation is achieved by the adjustment of the classical Galilean transformations resulting in

$$x = \gamma(x_o - wt_o), y = y_o, z = z_o, t = \gamma^{-1}t_o \qquad (1.3)$$

Here the zero-subscript coordinates are the coordinates of space and time in the preferred frame, the unsubscripted coordinates are coordinates in a reference frame moving at speed $w$ relative to the preferred frame and $\gamma$ is the FLL contraction factor given by

$$\gamma = (1 - w^2/c^2)^{-1/2} \qquad (1.4)$$

Many researchers [14-17] have observed that the Absolute Space Theory is in close agreement with Special Relativity Theory over virtually its full range of predictions.

However the two theories make quite different light speed predictions. In the case of the Absolute Space Theory, for measurements made by an observer at rest in the preferred frame, the (real) speed $u'_r$ relative to the moving frame is given by

$$u'_r = u - w \qquad (1.5)$$

where $u$ is the speed relative to the preferred frame. This is the Galilean law of velocity composition. For measurements made by an observer moving relative to the preferred frame, Levy [16 p42-43] has shown that because the FLL contractions result in contracted metre sticks and retarded clocks, the (apparent) speed $u'_a$ relative to the moving frame is given by

$$u'_a = (u - w)/(1 - w^2/c^2) \qquad (1.6)$$

which can be written as

$$u'_a(1 - w^2/c^2) = u - w \qquad (1.7)$$

This is the Galilean law of velocity composition when contracted metre sticks and retarded clocks are used to measure speed relative to the moving frame. From (1.5) and (1.7), the real speed $u'_r$ and the apparent speed $u'_a$ are related by

$$u'_a(1 - w^2/c^2) = u'_r \qquad (1.8)$$

The law of velocity composition in Special Relativity Theory corresponding to equation (1.6) of the Absolute Space Theory is [1-4]

$$u' = (u - w)/(1 - uw/c^2) \tag{1.9}$$

where $u'$ is the speed relative to the moving frame. If $u = c$ in (1.9), then

$$u' = c \tag{1.10}$$

i.e. the measured speed in Special Relativity Theory is $c$. This result is independent of the direction of $w$ and consistent with the light speed invariance postulate. If $u = c$ in (1.6) of the Absolute Space Theory, then

$$u'_a = (c - w)/(1 - w^2/c^2) \approx c - w, w << c \tag{1.11}$$

i.e. for sufficiently low relative speed $w$, the measured speed in Absolute Space Theory is the classical value $c - w$ and not $c$. If the direction of $w$ is reversed, then (1.11) becomes

$$u'_a = (c + w)/(1 - w^2/c^2) \approx c + w, w << c \tag{1.12}$$

Thus for light speed measurements made by an observer in a frame moving at speed $w$ relative to the preferred frame, Special Relativity Theory predicts light speed invariance $c$ while the Absolute Space Theory predicts classical light speed variation $c \pm w$. In this paper therefore, we examine the possibility of light speed variation and its measurability in the framework of the Absolute Space Theory. To this end, we summarise the careful results of Ives relating to out-and-back optical tests which show how experiments such as the Michelson-Morley and Kennedy-Thorndike involving light speed variation in a preferred frame yield a measured constant light speed $c$. We then show that light speed variation can indeed be detected. Specifically we demonstrate that the variation in the period of Jupiter's satellite Io observed from Earth as it orbits the Sun ("Roemer Effect") is a direct manifestation of changes in the speed of light relative to a moving observer.

## 2. Light Speed Measurement on a Moving Platform

Consider the Maxwell-Lorentz semi-classical Absolute Space Theory in which light propagates isotropically at a speed $c$ in a preferred frame. In such an absolute frame, the one-way speed of light relative to an observer changes according to the observer's motion relative to the preferred frame. In addition, FLL contractions (1.1) and (1.2) occur as a result of movement relative to the preferred frame. These contractions alter the normal Galilean transformations that relate the coordinates of the preferred frame to coordinates in any other inertial frame, which now become (1.3).

## 2.1 Two-way Light Speed Test [18]

Since one-way light speed methods have failed because of clock synchronization problems [5], we here focus on two-way light speed testing which appears to confirm light speed $c$. We consider light speed measurement on a platform moving at a speed $w$ relative to the postulated preferred reference frame and examine the possibility of detecting and measuring the speed $c - w$ of similarly directed light relative to the platform [18]. In Figure 1, let ab be the moving platform on which measurements are to be made with a clock at a and a mirror at b. Let the speed of the platform relative to the preferred reference frame be $w$ and let $\overline{D}$ be the distance from a to b as measured on the platform.

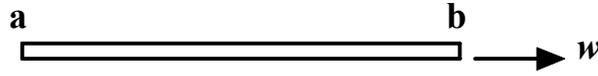

**Figure 1. Platform moving at speed $w$ relative to the preferred reference frame**

Because of the FLL contractions that arise as a result of movement with respect to the preferred frame, the true length $D$ is less than the measured length and is given by

$$D = \overline{D}(1 - w^2 / c^2)^{1/2} \tag{2.1}$$

Let the time of transit of a light signal travelling from a to b and back to a as measured by the clock at a be $\bar{t}$. Because of the FLL contractions, the true time of transit $t$ (measured by a clock stationary in the preferred frame) from a to b and back is greater than $\bar{t}$ and given by

$$t = \frac{\bar{t}}{(1 - w^2 / c^2)^{1/2}} \tag{2.2}$$

Now the true time of transit of a light signal from a to b is $D/(c - w)$ and the true time of transit from b to a is $D/(c + w)$. Therefore

$$t = \frac{D}{c - w} + \frac{D}{c + w} = \frac{2cD}{c^2 - w^2} \tag{2.3}$$

Substituting $D$ from (2.1) and $t$ from (2.2) in (2.3) yields

$$\frac{2c\overline{D}}{c^2 - w^2}(1 - w^2 / c^2)^{1/2} = \frac{\bar{t}}{(1 - w^2 / c^2)^{1/2}} \tag{2.4}$$

It follows from (2.4) that the speed of light $\bar{c}$ as measured on the moving platform is given by

$$\bar{c} = \frac{2\bar{D}}{\bar{t}} = c \tag{2.5}$$

Therefore even though the speed of light relative to the platform is $c - w$ out and $c + w$ back, the light speed measured on the platform is $c$.

To appreciate the full significance of this remarkable result, it should be noted that the true average out and back speed $c_{AV}$ is, using (2.3), calculated to be

$$c_{AV} = \frac{2D}{t} = \frac{2D}{2cD}(c^2 - w^2) = c(1 - w^2 / c^2) \neq c \tag{2.6}$$

The result in (2.6) indicates that with no FLL contractions, the average two-way light speed $c_{AV}$ varies with $w$ to second order and is not equal to $c$: The FLL contractions compensate for the second-order term $w^2 / c^2$ in (2.6) such that the measured average two-way speed $\bar{c}$ is $c$ as given in (2.5). This result has been generalized for any direction of light travel by Levy [16] and validates the standard out-and-back method of determining light speed $c$ relative to the preferred frame from a moving platform. It means however that this method gives $c$ relative to the preferred frame but does not yield the one-way light speed $c \pm w$ relative to the moving platform.

## 3. Light Speed Measurement Using a One-way Signal Pulse Train

Neither one-way two-clock light speed experiments nor two-way one-clock light speed experiments give the one-way light speed relative to the moving platform. We now describe a one-way one-clock experiment that does. It is based on the following principle: Instead of timing a one-way light signal pulse over a known distance using two clocks, we time successive pulses of a one-way signal train using one clock. It is somewhat similar to timing a passenger train with one clock by starting the clock as the front of the train passes and stopping the clock as the rear of the train passes. If the length $D_T$ of the train is known, then with the time $\overline{T_T}$ measured by the single clock, the train speed $\overline{S_T}$ can be found from

$$\overline{S_T} = D_T / \overline{T_T} \tag{3.1}$$

### 3.1 The Roemer Experiment

Consider the Earth-Sun-Jupiter planetary system. As the Earth revolves around the Sun at speed $w$, the innermost satellite of Jupiter, Io, is observed to undergo regular variations in its period $T_o$ [19]. Because Io, as observed from Earth, is periodically eclipsed by Jupiter, this occulting source emits what may be described as "pulses of darkness" travelling at speed $c$ to Earth as Io revolves around Jupiter. This is shown in figure 2. The distance between successive pulses is fixed at $\lambda_o$ given by

$$\lambda_o = cT_o \qquad (3.2)$$

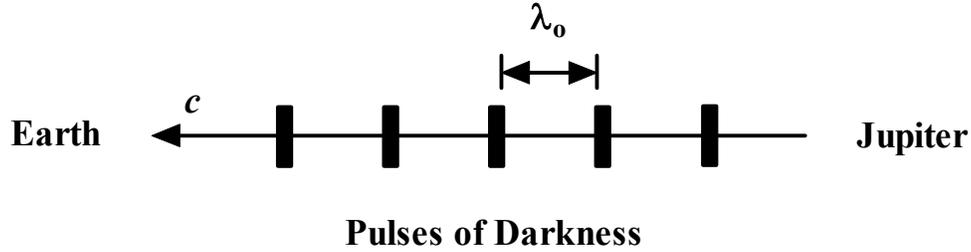

**Figure 2. Pulses of Darkness from Earth-Sun-Jupiter Planetary System**

As the Earth moves away from Jupiter, the period $\overline{T_H}$ of Io, which is the time between successive pulses, is observed on Earth and found to be greater than $T_o$. The light speed relative to the receding Earth can now be determined using $\lambda_o$ and $\overline{T_H}$ in the fundamental speed-determining equation "speed equals distance over time" which for this system is

$$\overline{c_R} = \frac{\lambda_o}{\overline{T_H}} \qquad (3.3)$$

Of course, as in the case of the train, in order to determine the actual speed, the distance $\lambda_o$ must be known. This is easily obtained by measuring $\overline{T_H}$ for $w = 0$ corresponding to $T_o$ (by for example employing a space probe), measuring $c$ using the independent out-and-back method, and using these values in (3.2) to determine $\lambda_o$. Establishing the value of $\lambda_o$ using $\overline{T_H}(w=0)$ and an independently determined $c$ may be viewed as calibration of the measurement apparatus; once this is done, all light speed variations relative to the receding Earth can then be directly determined by measuring $\overline{T_H}$ and using (3.3).

In order to verify this method of determining variable light speed, we determine light speed relative to the receding Earth by the direct substitution of measured values for $T_o, \overline{T_H}$ and $c$ in the equation (3.3) for light speed. Thus using (3.2) in (3.3) yields

$$\overline{c_R} = \frac{\lambda_o}{T_H} = \frac{cT_o}{T_H} \tag{3.4}$$

Substituting in (3.4) the measured values $T_o = 152{,}944s$ and $\overline{T_H} = 152{,}959.2s$ corresponding to the increased period of Io [19] along with the independently determined $c = 299{,}792{,}458ms^{-1}$ gives the relative light speed value

$$\overline{c_R} = \frac{299{,}792{,}458 \times 152{,}944}{152{,}959.2} = 299{,}762{,}667ms^{-1} \tag{3.5}$$

Note that since $w \ll c$, second-order FLL contraction effects associated with the measurement of $\overline{T_H}$ are small and therefore do not significantly affect this result. We observe that the experimentally determined light speed value in (3.5) is not $c$ as the light speed invariance postulate demands. The light speed in (3.5) is almost exactly equal to the classical value of relative light speed value $c - w$ for the receding Earth, which using $w = 29{,}790ms^{-1}$ for the Earth is

$$c - w = (299{,}792{,}458 - 29{,}790)ms^{-1} = 299{,}762{,}668ms^{-1} \tag{3.6}$$

Therefore for movement of the Earth directly away from Jupiter, the relative light speed $\overline{c_R}$ is given by

$$\overline{c_R} = \frac{\lambda_o}{T_H} = c - w \tag{3.7}$$

For movement of the Earth directly towards Jupiter, the period $\overline{T_L}$ of Io (which again is the time between successive pulses) is observed on Earth and found to be less than $T_o$. The light speed relative to the advancing Earth can be determined using $\lambda_o$ and $\overline{T_L}$ in the fundamental speed equation for this system which is

$$\overline{c_R} = \frac{\lambda_o}{T_L} \tag{3.8}$$

Substituting in (3.8) the measured values $T_o = 152{,}944s$ and $\overline{T_L} = 152{,}928.8s$ corresponding to the shortened period of Io [19] along with $c = 299{,}792{,}458ms^{-1}$ gives the relative light speed value

$$\overline{c_R} = \frac{299{,}792{,}458 \times 152{,}944}{152{,}928.8} = 299{,}822{,}255ms^{-1} \tag{3.9}$$

We again observe that the experimentally determined light speed value in (3.9) is not $c$ as the light speed invariance postulate requires. The light speed in (3.9) is almost exactly equal to the classical value of relative light speed $c + w$ for the advancing Earth, which using $w = 29{,}790 ms^{-1}$ for the Earth is

$$c + w = (299{,}792{,}458 + 29{,}790) ms^{-1} = 299{,}822{,}248 ms^{-1} \tag{3.10}$$

Therefore for movement of the Earth directly towards Jupiter, the relative light speed $\overline{c_R}$ is given by

$$\overline{c_R} = \frac{\lambda_o}{\overline{T_L}} = c + w \tag{3.11}$$

On the basis of the experimentally demonstrated classical light speed variations in (3.7) and (3.11) relative to the moving Earth, we conclude that the change in the period of the planetary satellite Io measured by an observer on the Earth, is a direct indication of a change in light speed relative to that moving observer. The results (3.7) and (3.11) directly confirm the light speed predictions (1.11) and (1.12) of the Absolute Space Theory and falsify the light speed invariance postulate of Special Relativity Theory. Additionally, these light speed variations represent detection of movement of the Earth relative to the ether in the Earth's approximately uniform motion around the Sun, exactly the motion that Michelson and Morley failed to detect also using light speed variation in their unsuccessful second-order experiment of 1887 [6].

## 4.   Conclusion

The light speed invariance postulate that underpins Special Relativity Theory has not been directly confirmed for one-way light transmission. For two-way light travel, the one-clock out-and-back measurement of light speed on a moving platform always yields the value $c$ and this appears to confirm the postulate. This is reinforced by the failure to detect light speed anisotropy in several one-way two-clock tests [10, 11]. However, these one-way two-clock tests yielding $c$ are flawed because of clock synchronization problems [5, 16]. Further, as shown in this paper, the light speed invariance observed in the two-way one-clock measurement is an ***illusion*** created by the compensating effect of the FLL contractions and consequent elimination of platform speed $w$ from the measured speed. By using a one-way one-clock measurement procedure, we escaped the compensating effect of the FLL contractions as well as the clock synchronization problems and thereby succeeded in revealing variable light speed $c \pm w$ relative to a moving frame.

We demonstrated that in the Roemer experiment involving the moving Earth and an occulting light source Io, changes in the speed of light relative to the moving Earth $c \pm w$ do occur and are measurable and result in observable variations in the period of Io as seen on Earth. The light speed variation established in the Roemer experiment directly contradicts Einstein's Light Speed Invariance Postulate. Therefore this postulate, which results from what Will ([20],p247) referred to as "Einstein's revolutionary insight" and which Kaku ([21],p82) described as "one of the greatest achievements of the human spirit", is wrong. As a result Special Relativity Theory, which is based on this postulate and which also predicts it, collapses [22]!

In addition to the falsification of the light speed invariance postulate, the measured light speed anisotropy in the Roemer experiment represents detection of ether drift and confirms the existence of the preferred frame of the Absolute Space Theory [16, 17] in which light propagates at speed $c$. This detection of the movement of the Earth relative to this preferred frame or ether in the Earth's approximately uniform motion around the Sun was recently reported by Gift [23] based on light speed variation in the Roemer and Doppler experiments. The existence of this preferred frame is consistent with the absolute speed measurement of the Earth arising from anisotropic measurement of the cosmic microwave background radiation [24] and the apparent determination of absolute motion in experiments by Marinov [25] and Silvertooth [26].

In light of the incontrovertible revelation in this paper of light speed anisotropy occurring in the physical world and the consequent invalidation of Special Relativity Theory, we urge the scientific community to reject the century-old relativistic doctrine of Albert Einstein with its many paradoxes and contradictions and return to the absolute frame of the causal ether embodied in the space-time framework of the modern Maxwell-Lorentz Absolute Space Theory [16, 17]. The luminiferous ether, whose existence was accepted by all scientists up to the end of the 19[th] century but later abandoned because of the failure of the 1887 experiment to detect it, is real and should now be the focus of exhaustive scientific investigation.